\documentclass[epj]{svjour}
\usepackage{graphicx}
\usepackage{amsmath}
\usepackage{cuted}
\usepackage{url}
\graphicspath{{./figures/}}
\newcommand{\Tr}{\mathop{\mathrm{Tr}}\nolimits}
\renewcommand{\Re}{\mathop{\mathrm{Re}}\nolimits}
\begin{document}
\title{Finite temperature QCD with
$N_f=2+1+1$ Wilson twisted mass fermions at
physical pion, strange and charm masses}
\titlerunning{Finite T QCD with $N_f=2+1+1$ Wilson fermions}
\author{Andrey Yu. Kotov\inst{1}\thanks{andrey.kotov@phystech.edu}, Maria Paola Lombardo\inst{2} \and Anton M. Trunin\inst{3}
}
\authorrunning{A.Yu.~Kotov, M.P.~Lombardo, A.M.~Trunin}
\institute{Moscow Institute of Physics and Technology, Institutsky lane 9, Dolgoprudny, Moscow region, 141700 Russia\\
National University of Science and Technology MISIS, Leninsky Prospect 4, Moscow, 119049 Russia\\
Bogoliubov Laboratory of Theoretical Physics, Joint Institute for Nuclear Research, Dubna, 141980 Russia
\and INFN, Sezione di Firenze, 50019 Sesto Fiorentino (FI), Italy
\and Samara National Research University, Samara, 443086 Russia}
\date{Received: date / Revised version: date}
\abstract{
We discuss recent progress in studying Quantum Chromodynamics at finite temperature using $N_f=2+1+1$ Wilson twisted mass fermions. Particular interest is in QCD symmetries and their breaking and restoration. First, we discuss the behaviour of the $\eta'$ meson at finite temperature, which is tightly connected to the axial and chiral symmetries. The results suggest a small decrease of the $\eta'$ mass in the pseudo-critical region coming close to the non-anomalous contribution and subsequent growth at large temperatures. Second, we present the first results of lattice simulations of Quantum Chromodynamics with $N_f=2+1+1$ twisted mass Wilson fermions at physical pion, strange and charm masses. We estimate the chiral pseudo-critical temperatures for different observables. Our preliminary results are consistent with a second order transition in the chiral limit, however other scenarios are not excluded. 
} 

\maketitle

\section{Introduction}
\label{intro}

Quantum Chromodynamics under extreme conditions has been the subject of numerous theoretical and experimental studies\cite{Busza:2018rrf}. In the experiments on heavy ion collisions at Relativistic Heavy Ion Collider, Brookhaven and Large Hadron Collider, CERN
a droplet of strongly coupled matter at large temperatures is believed to be produced, thus providing a great opportunity to study thermal QCD. One of the most famous experimental discoveries was the observation of the Quark-Gluon Plasma
~--- a new state of matter, characterized by unbound deconfined quarks and gluons. From the theoretical side, the existing information on finite temperature QCD is based on  first-principle lattice supercomputer simulations, see e.g. Ref. \cite{Ding:2020rtq} for a recent review.

Properties of strongly interacting matter at nonzero temperature are tightly related to the symmetries and symmetry breaking pattern of QCD\cite{Philipsen:2019rjq}. The chiral symmetry $SU_L(2)\times SU_R(2)$, being broken in the vacuum state of QCD, becomes effectively restored at temperature $T_c\approx160$ MeV\cite{Borsanyi:2016ksw,Bazavov:2018mes,Steinbrecher:2018phh}. Approximately at the same temperature the transition to the deconfined phase of Quark-Gluon Plasma occurs with both transitions being analytical crossovers, rather than the phase transitions\cite{Aoki:2006we}.

 The behaviour of $U_A(1)$ axial symmetry at finite temperature is a more subtle issue. The proposed mechanism of instanton suppression might lead to the effective restoration of the axial symmetry close to the chiral (pseudo-)critical temperature\cite{Shuryak:1993ee}. Numerical lattice studies of various observables with different fermion discretisations give rather controversial results --- some propose joint effective restoration of axial and chiral symmetry,  while others favour axial symmetry restoration at much higher temperatures (see \cite{Philipsen:2019rjq} and references therein).
 From a phenomenological point of view, the restoration of axial symmetry should be reflected in the particle spectrum as a degeneracy of axial partners, see \cite{Kotov:2019dby}
 and references therein. 
 
 The behaviour of $U_A(1)$ symmetry also has a clear link to the universality class of the finite temperature QCD phase transition\cite{Pisarski:1983ms,Pelissetto:2013hqa}. If it remains broken after the chiral phase transition, the transition in the chiral limit should be in the $O(4)$-universality class. Effective restoration of $U_A(1)$ implies enlarged symmetry breaking pattern and, consequently, another behaviour of the chiral transition: it should be either first-order or in the other universality class $U(2)\otimes U(2)/U(2)$. The scaling of chiral observables with the quark mass could in principle distinguish possible scenarios, thus calling for simulations at low pion mass, even lower than physical\cite{Ding:2019prx}.

Apart from axial and chiral  symmetry restoration, the existence of other thresholds in Quark-Gluon Plasma was proposed\cite{Rohrhofer:2019qwq,Alexandru:2019gdm} suggesting possible emergence of more elaborate symmetries. So far, the thermal QCD and its symmetries and symmetry breaking patterns are not yet fully understood and further first-principle results are required.

In this note we present lattice results for different pion
masses, including, for the first time, the physical mass.
The gauge field configurations were generated using the public ETMC code. Our setup for pion masses ranging from 210 MeV till 470 MeV is taken from Ref. \cite{Carrasco:2014cwa},  
while for the physical pion mass we use the recent tuning
of Refs. \cite{Alexandrou:2018egz,Dimopoulos:2020eqd}.

\begin{figure}
    \centering
    \includegraphics[width=0.5\textwidth]{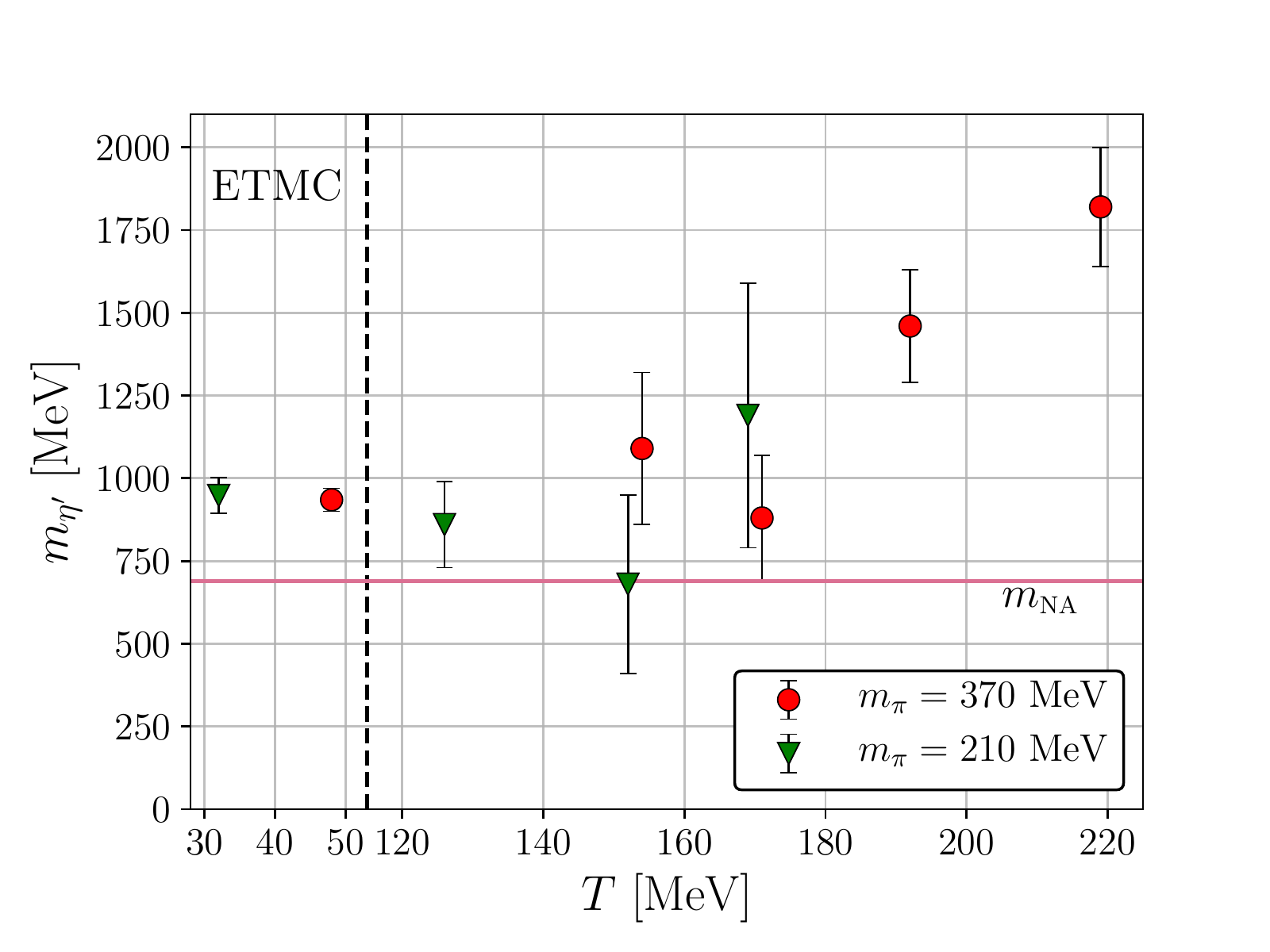}
    \caption{The dependence of the $\eta'$ mass on the temperature for pion masses $m_{\pi}=210$ MeV and $370$ MeV. The data for low temperatures are taken from \cite{Ottnad:2017bjt}. Purple horizontal line corresponds to non-anomalous contribution $m_{\text{NA}}\approx700$ MeV. From Ref.~\cite{Kotov:2019dby}}
    \label{fig:etaprimemass}
\end{figure}

\section{$\eta'$ and QCD symmetries}
\label{sec:eta}

The behaviour of the axial and chiral symmetries in Quantum Chromodynamics is tightly connected with the meson spectrum. Spontaneous breaking of chiral symmetry leads to an octet of Goldstone bosons: $\pi$'s, $K$'s and $\eta$. Nonzero quark masses $m_q$, breaking chiral symmetry explicitly, lead to small nonzero masses of these pseudo-Goldstone bosons $M_{PGB}^2\sim m_q \Lambda_{QCD}$. If $U_A(1)$ axial symmetry would be also spontaneously broken, its would-be Goldstone boson $\eta'$ mass should follow the same pattern. However, the experimental result for $m_{\eta'}\sim 1$ GeV is much higher than the masses of the octet of pseudoscalar mesons. The solution to this  well known puzzle comes from the fact that the axial symmetry is anomalously broken and $\eta'$ mass gets an additional contribution coming from axial anomaly\cite{tHooft:1976rip}. This contribution can be quantitatively taken into account within large-$N$ limit\cite{Veneziano:1979ec}. There are studies of the $\eta/\eta'$-complex beyond large-$N$ approximation, both phenomenological\cite{Shore:2007yn} and on the lattice\cite{Kaneko:2009za,Christ:2010dd,Gregory:2011sg,Ottnad:2015hva,Fukaya:2015ara}. In general, the behaviour of the $\eta'$  in QCD at zero temperature is well understood (see, e.g., the detailed discussion in \cite{Kotov:2019dby}). The properties of $\eta'$ at finite temperature are less clear and require first-principle investigation.

 To determine the $\eta'$-mass at nonzero temperature we measured the correlator of the topological charge density $G^q(t)=\int d^3\bar{x}\langle q(\bar{x},t)q(0,0)\rangle$, which is coupled to $\eta'$ due to the axial anomaly. 
Topological charge density was measured with the help of Gradient Flow\cite{Luscher:2010iy}. By fitting the data at large Euclidean times with a simple behaviour $G(\tau)\sim\cosh\left[m(\tau-N_{\tau}/2)\right]$, where $N_{\tau}$ is the temporal lattice extent, we extracted the parameter $m$, which corresponds to the $\eta'$ mass. More detailed description of the simulation setup can be found
in our original paper\cite{Kotov:2019dby}. The final results for the temperature dependence of the $\eta'$-mass for two pion masses $m_{\pi}=370$ MeV and $m_{\pi}=210$ MeV are presented in Fig.~\ref{fig:etaprimemass}.

The results suggest a small decrease in the $\eta'$ mass in the vicinity of the chiral pseudo-critical temperature followed by an increase at larger temperatures. Various phenomenological studies\cite{Horvatic:2018ztu,Nicola:2018vug,Nicola:2019ohb,Ishii:2016dln,Mitter:2013fxa} provide similar trend, although some quantitative features are different. One of the possible explanation of these results would be the restoration of chiral and axial symmetries, so that the whole nonet of pseudo-Goldstone bosons becomes degenerate. To draw more definite conclusions, one needs more information from other observables, sensitive to chiral and axial symmetry.
In particular, it is very important to measure the spectrum of other particles, both already mentioned pseudo-Goldstone bosons as well as other mesons, at finite temperature, and to
extend the results to physical pion mass. This is the subject of an ongoing study. 

\section{Chiral phase transition for physical pion mass}
\label{sec:physpionmass}

One of the challenges of modern lattice studies of QCD is related to the fact that the required computational time grows very fast with the pion mass going down to its physical value.
So far, the main results at the physical pion mass $m_{\pi}\sim140$ MeV were obtained for (improved) staggered fermions\cite{Borsanyi:2016ksw,Bazavov:2018mes}, although results with other discretizations are also available\cite{Bhattacharya:2014ara}. One of the alternatives is to use
Wilson-type fermions, in particular, twisted-mass implementation of Wilson fermions\cite{Frezzotti:2000nk}. Development of  multi-grid algorithm\cite{Alexandrou:2016izb} made it possible to perform simulations with twisted mass fermions at physical pion mass. 

Our setup is based on recent tuning of the parameters by ETM collaboration at $T=0$ for physical pion mass\cite{Alexandrou:2018egz,Dimopoulos:2020eqd}.
Following Refs.~\cite{Alexandrou:2018egz,Dimopoulos:2020eqd}, 
simulations were performed with $N_f=2+1+1$ Wilson  twisted mass fermions at the isospin symmetric point. With respect to our
previous study, a clover term was included, and 
the fermionic action for two light quarks $S^{l}$ and for $1+1$ heavy doublet $S^{h}$  has the following form:

\begin{strip}
\begin{equation}
\begin{split}
    S^{l}=\sum_{x,y}\bar{\chi}_l(x)\left[\left(1-\frac{i}{2}\kappa c_{SW}\sigma^{\mu\nu}\mathcal{F}^{\mu\nu}\right)\delta_{x,y}-\kappa D_W[U](x,y)+2i\kappa a\mu_l\tau^3\gamma^5\delta_{x,y}\right]\chi_l(y)\\
    S^{h}=\sum_{x,y}\bar{\chi}_h(x)\left[\left(1-\frac{i}{2}\kappa c_{SW}\sigma^{\mu\nu}\mathcal{F}^{\mu\nu}\right)\delta_{x,y}-\kappa D_W[U](x,y)+2i\kappa a\mu_{\sigma}\tau^1\gamma^5\delta_{x,y}+2\kappa a\mu_{\delta}\tau^3\delta_{x,y}\right]\chi_h(y),
\label{eq:fermionaction}
\end{split}
\end{equation}
\end{strip}
where $a$ is the lattice spacing, $D_W[U]$ is the usual Wilson operator, $c_{SW}\sigma^{\mu\nu}\mathcal{F}^{\mu\nu}$ is the standard clover term\cite{Hasenbusch:2002ai}. Parameters of the action were tuned by zero temperature simulations of ETM collaboration\cite{Alexandrou:2018egz} to reproduce  physical pion mass $m_{\pi}=139.3(7)$~MeV\cite{Alexandrou:2018sjm}. 

For the gauge fields the Iwasaki improved action was used ($c_0=3.648$, $c_1=-0.331$):

\begin{strip}
\begin{equation}
S^g=\beta\sum (c_0\sum_P[1-\frac{1}{3}
\Re\Tr U_P]+c_1\sum_R[1-\frac13\Re\Tr U_R] ). 
\label{eq:gaugeaction}
\end{equation}
\end{strip}
Here $\sum\limits_P$ and $\sum\limits_R$ denote the sum over all $1\times1$ plaquettes and over all $1\times2$ rectangles, correspondingly. Inverse bare gauge coupling was fixed at $\beta=1.778$.

\begin{table}
\caption{Statistics used in the simulations}
\label{tab:statistics}       
\centering
\begin{tabular}{ccc}
\hline\noalign{\smallskip}
$N_t$ & $T$ [MeV] & \# conf  \\
\noalign{\smallskip}\hline\noalign{\smallskip}
20 & 123(1) & 244 \\
18 & 137(1) & 155 \\
16 & 154(1) & 364 \\
14 & 176(1) & 129 \\
12 & 205(1) & 263 \\
10 & 246(1) & 205 \\
8 & 308(2) & 460 \\
6 & 411(2) & 195 \\
4 & 616(3) & 472 \\
\noalign{\smallskip}\hline
\end{tabular}
\end{table}

Simulation are performed in a fixed-scale approach, where temperature $T=\frac{1}{N_ta}$ is varied by varying the temporal extent of the lattice $N_t$, and  all the parameters of the action (\ref{eq:fermionaction})--(\ref{eq:gaugeaction}) are kept constant. We present the results for one lattice spacing $a=0.0801(4)$ fm\cite{Alexandrou:2018sjm}. Summary of all used statistics for various values of $N_t$ is presented in Tab.~\ref{tab:statistics}.
Simulations are still in progress, and complete results will be reported elsewhere. 

\begin{figure}
    \centering
    \includegraphics[width=0.5\textwidth]{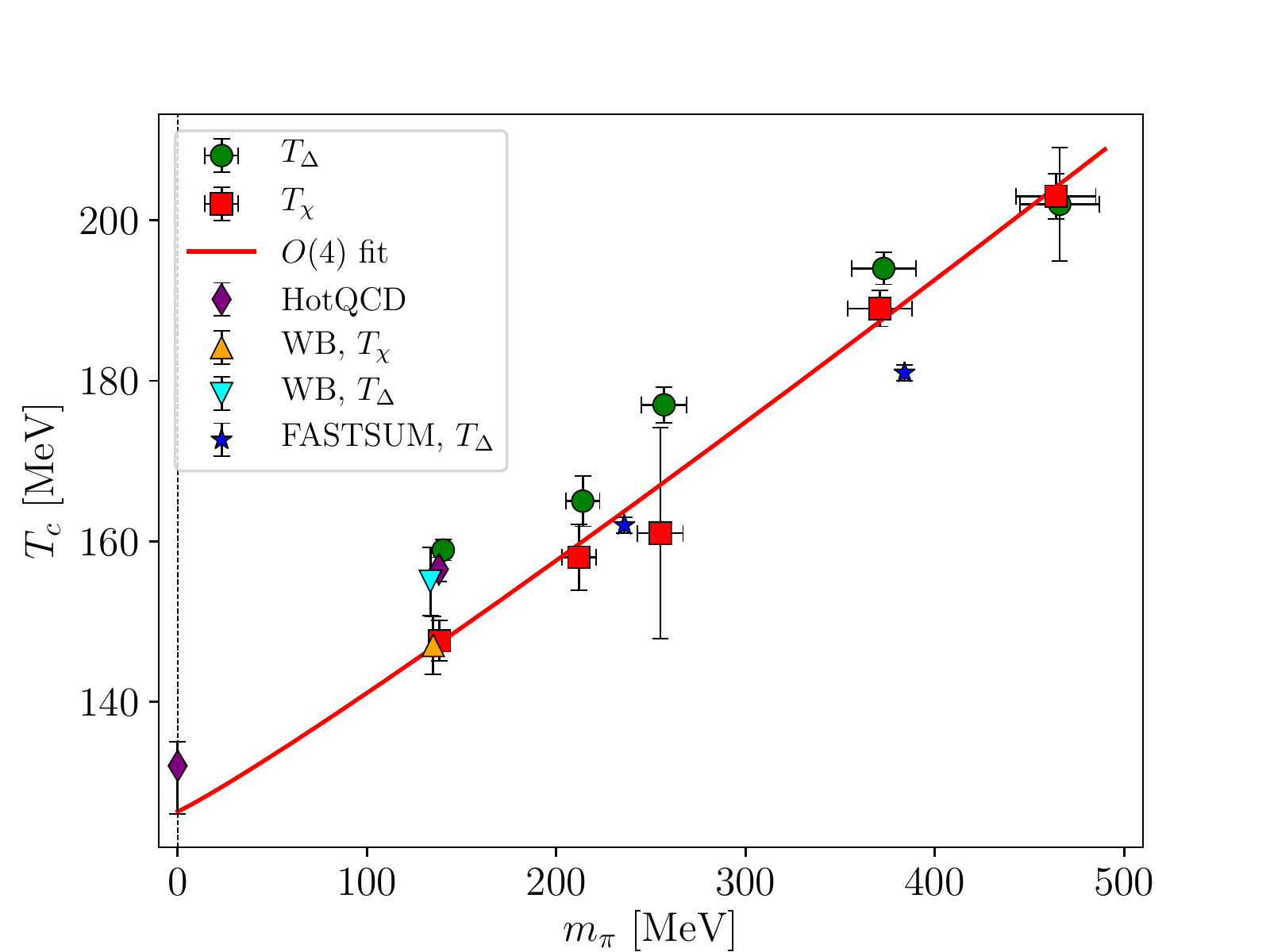}
    \caption{Pseudocritical temperature determined from the inflection point of the chiral condensate $T_{\Delta}$ (green circles) and peak of the susceptibility $T_{\chi}$ (red squares) versus the pion mass. The data for higher than physical pion masses are taken from \cite{Burger:2018fvb}. Along with our results we present the staggered results of Wuppertal--Budapest collaboration (cyan and orange triangles)\cite{Borsanyi:2010bp}, recent results of HotQCD collaboration both at physical pion mass and in the chiral limit (purple rhombi)\cite{Bazavov:2018mes} and preliminary results of the FASTSUM collaboration (blue stars)\cite{Aarts:2019hrg}. Red line corresponds to the fit $T_{\chi}\propto T_c+Am_{\pi}^x$ with $x$  fixed to $O(4)$ value $x=1.083$. 
    }
    \label{fig:crit_temp}
\end{figure}

In our first study with $N_f=2+1+1$ twisted mass fermions at physical pion mass we measured simple chiral observables: the light quark chiral condensate $\langle \bar{\psi}\psi\rangle_l$ and the disconnected chiral susceptibility $\chi^{\textrm{disc}}_{\bar{\psi}\psi}=\frac{V}{T}\bigl(\langle(\bar{\psi}\psi)^2\rangle_l-\langle\bar{\psi}\psi\rangle_l^2\bigr)$.
Using the dependence of these observables on temperature we estimated the pseudocritical temperature of the chiral phase transition, as the inflection point of the chiral condensate $\langle \bar{\psi}\psi\rangle_l$ versus $T$ and the peak in the susceptibility $\chi^{\textrm{disc}}_{\bar{\psi}\psi}$. To extract the inflection point $T_{\Delta}$, the chiral condensate in the transition region was fitted by several functions $\langle \bar{\psi}\psi\rangle=A+B\tanh{\frac{T-T_{\Delta}}{\delta T_{\Delta}}}$, $\langle \bar{\psi}\psi\rangle=A+B(T-T_{\Delta})/\sqrt{\delta T^2+(T-T_{\Delta})^2}$ and $\langle \bar{\psi}\psi\rangle=a_{\Delta}+b_{\Delta}T+c_{\Delta}T^2+d_{\Delta}T^3$. The difference of $T_{\Delta}$ extracted from various fits, and by varying fitting interval, allowed us to estimate the systematic uncertainty of our results.
In the same way by using various functions $\chi^{\textrm{disc}}_{\bar{\psi}\psi}=A_0+B_0(T-T_{\chi})^2$, $\chi^{\textrm{disc}}_{\bar{\psi}\psi}=A/(B^2+(T-T_{\chi})^2)$ and $\chi^{\textrm{disc}}_{\bar{\psi}\psi}=Ae^{-\frac{(T-T_c)^2}{\delta T^2}}$ and various fitting ranges we estimated the peak of the chiral susceptibility $T_{\chi}$. 

In Fig.~\ref{fig:crit_temp} we present the results for the pseudocritical temperature extracted from the inflection point of the chiral condensate $T_{\Delta}$ and the peak of the chiral susceptibility $T_{\chi}$. The data for higher than physical pion masses are taken from \cite{Burger:2018fvb}.
In the same plot we show the results for the physical pion mass obtained with staggered fermions\cite{Bazavov:2018mes,Borsanyi:2010bp}, as well as preliminary results for heavier masses from Wilson fermions on asymmetric
lattices from the FASTSUM collaboration\cite{Aarts:2019hrg}.
Preliminary fit of $T_\chi$ to a power law behavior 
$T_\chi(m_\pi) \propto T_c + A m_\pi^x $, with $x$  constrained to the $O(4)$ value $x = 2/{\beta \delta} \simeq  1.083$, is superimposed. A fit 
corresponding to a critical behaviour in the 
$Z_2$ universality class  associated with an endpoint of
a hypothesized first order phase transition
at $m_{crit} < m_\pi$ would
also describe the data.  We postpone a discussion of these different scenarios to a lengthier publication\cite{progress}. 

\section{Discussion and further directions}
\label{discussion}

Recent progress in algorithms and supercomputers leads to significant advance in first-principle lattice simulations. In this work we have presented the first results for a physical pion mass with twisted mass Wilson fermions. Our first results for the pseudo-critical temperature with $N_f=2+1+1$ twisted mass fermions at physical pion mass are consistent with an $O(4)$ universality
class as found with  staggered fermions\cite{Steinbrecher:2018phh}.

However, the properties of Quantum Chromodynamics at nonzero temperature still are not yet fully understood. The properties of the chiral transition, its behaviour and universality class in the chiral limit, and the role of the axial symmetry near the chiral phase transition remain an open issue and require further attention. Although our results for the $\eta'$ mass are in favour of disappearance of axial anomaly and restoration of the axial symmetry in the vicinity of the chiral phase transition, a more thorough study of various observables is required to draw more definite conclusions. In particular, in the future, alongside with a more complete analysis of the scaling properties of the order parameters, a detailed study of meson spectrum for physical pion mass is planned\cite{progress}.

\acknowledgement

It is a pleasure to thank Roberto Frezzotti for useful 
conversation on twisted mass Wilson fermions. 
A.M.T. acknowledges support from the ``BASIS" foundation. A.Yu.K. acknowledges the hospitality of the Galileo Galilei Institute for Theoretical Physics and the support of the European COST Action CA15213 ``Theory of hot matter and relativistic heavy-ion collisions" (THOR). The work of A.Yu.K. was also supported by RFBR grant 18-02-40126.
Numerical simulations have been carried out using computing resources of CINECA (based on the agreement between INFN and CINECA, on the ISCRA project IsB20), the supercomputer of Joint Institute for Nuclear Research ``Govorun'' and the computing resources of the federal collective usage center Complex for Simulation and Data Processing for Mega-science Facilities at NRC ``Kurchatov Institute'',~\url{http://ckp.nrcki.ru/}.


\begin{thebibliography}{10}

\bibitem{Busza:2018rrf}
W.~Busza, K.~Rajagopal, and W.~van~der Schee.
\newblock {Heavy Ion Collisions: The Big Picture, and the Big Questions}.
\newblock {\em Ann. Rev. Nucl. Part. Sci.}, 68:339--376, 2018.

\bibitem{Ding:2020rtq}
H.-T. Ding.
\newblock {New developments in lattice QCD on equilibrium physics and phase
  diagram}.
\newblock In {\em {28th International Conference on Ultrarelativistic
  Nucleus-Nucleus Collisions}}, 2 2020.

\bibitem{Philipsen:2019rjq}
O.~Philipsen.
\newblock {Constraining the QCD phase diagram at finite temperature and
  density}.
\newblock In {\em {37th International Symposium on Lattice Field Theory
  (Lattice 2019) Wuhan, Hubei, China, June 16-22, 2019}}, 2019.

\bibitem{Borsanyi:2016ksw}
Sz. Borsanyi et~al.
\newblock {Calculation of the axion mass based on high-temperature lattice
  quantum chromodynamics}.
\newblock {\em Nature}, 539(7627):69--71, 2016.

\bibitem{Bazavov:2018mes}
A.~Bazavov et~al.
\newblock {Chiral crossover in QCD at zero and non-zero chemical potentials}.
\newblock {\em Phys. Lett.}, B795:15--21, 2019.

\bibitem{Steinbrecher:2018phh}
P.~Steinbrecher.
\newblock {The QCD crossover at zero and non-zero baryon densities from Lattice
  QCD}.
\newblock {\em Nucl. Phys.}, A982:847--850, 2019.

\bibitem{Aoki:2006we}
Y.~Aoki et~al.
\newblock {The Order of the quantum chromodynamics transition predicted by the
  standard model of particle physics}.
\newblock {\em Nature}, 443:675--678, 2006.

\bibitem{Shuryak:1993ee}
E.~V. Shuryak.
\newblock {Which chiral symmetry is restored in hot QCD?}
\newblock {\em Comments Nucl. Part. Phys.}, 21(4):235--248, 1994.

\bibitem{Kotov:2019dby}
A.~{\relax Yu}. Kotov, M.~P. Lombardo, and A.~M. Trunin.
\newblock {Fate of the $\eta'$ in the quark gluon plasma}.
\newblock {\em Phys. Lett.}, B794:83--88, 2019.

\bibitem{Pisarski:1983ms}
R.~D. Pisarski and F.~Wilczek.
\newblock {Remarks on the Chiral Phase Transition in Chromodynamics}.
\newblock {\em Phys.\ Rev.\ D}, 29:338--341, 1984.

\bibitem{Pelissetto:2013hqa}
A.~Pelissetto and E.~Vicari.
\newblock {Relevance of the axial anomaly at the finite-temperature chiral
  transition in QCD}.
\newblock {\em Phys. Rev.}, D88(10):105018, 2013.

\bibitem{Ding:2019prx}
H.~T. Ding et~al.
\newblock {Chiral Phase Transition Temperature in ( 2+1 )-Flavor QCD}.
\newblock {\em Phys. Rev. Lett.}, 123(6):062002, 2019.

\bibitem{Rohrhofer:2019qwq}
C.~Rohrhofer et~al.
\newblock {Symmetries of spatial meson correlators in high temperature QCD}.
\newblock {\em Phys. Rev.}, D100(1):014502, 2019.

\bibitem{Alexandru:2019gdm}
A.~Alexandru and I.~Horváth.
\newblock {Possible New Phase of Thermal QCD}.
\newblock {\em Phys. Rev.}, D100(9):094507, 2019.

\bibitem{Carrasco:2014cwa}
N.~Carrasco et~al.
\newblock {Up, down, strange and charm quark masses with N$_f$ = 2+1+1 twisted
  mass lattice QCD}.
\newblock {\em Nucl.\ Phys.\ B}, 887:19--68, 2014.

\bibitem{Alexandrou:2018egz}
C.~Alexandrou et~al.
\newblock {Simulating twisted mass fermions at physical light, strange and
  charm quark masses}.
\newblock {\em Phys. Rev.}, D98(5):054518, 2018.

\bibitem{Dimopoulos:2020eqd}
G.~Bergner et~al.
\newblock {Quark masses and decay constants in $N_f=2+1+1$ isoQCD with Wilson
  clover twisted mass fermions}.
\newblock In {\em {37th International Symposium on Lattice Field Theory}}, 1
  2020.

\bibitem{Ottnad:2017bjt}
K.~Ottnad and C.~Urbach.
\newblock {Flavor-singlet meson decay constants from $N_f=2+1+1$ twisted mass
  lattice QCD}.
\newblock {\em Phys. Rev.}, D97(5):054508, 2018.

\bibitem{tHooft:1976rip}
G.~'t~Hooft.
\newblock {Symmetry Breaking Through Bell-Jackiw Anomalies}.
\newblock {\em Phys. Rev. Lett.}, 37:8--11, 1976.

\bibitem{Veneziano:1979ec}
G.~Veneziano.
\newblock {U(1) Without Instantons}.
\newblock {\em Nucl. Phys.}, B159:213--224, 1979.

\bibitem{Shore:2007yn}
G.~M. Shore.
\newblock {The U(1)(A) Anomaly and QCD Phenomenology}.
\newblock {\em Lect. Notes Phys.}, 737:235--288, 2008.

\bibitem{Kaneko:2009za}
T.~Kaneko et~al.
\newblock {Flavor-singlet mesons in N(f) = 2+1 QCD with dynamical overlap
  quarks}.
\newblock {\em PoS}, LAT2009:107, 2009.

\bibitem{Christ:2010dd}
N.~H. Christ et~al.
\newblock {The $\eta$ and $\eta^\prime$ mesons from Lattice QCD}.
\newblock {\em Phys. Rev. Lett.}, 105:241601, 2010.

\bibitem{Gregory:2011sg}
E.~B. Gregory et~al.
\newblock {A study of the eta and eta' mesons with improved staggered
  fermions}.
\newblock {\em Phys. Rev.}, D86:014504, 2012.

\bibitem{Ottnad:2015hva}
K.~Ottnad, C.~Urbach, and F.~Zimmermann.
\newblock {A mixed action analysis of $\eta$ and $\eta'$ mesons}.
\newblock {\em Nucl. Phys.}, B896:470--492, 2015.

\bibitem{Fukaya:2015ara}
H.~Fukaya et~al.
\newblock {$\eta^\prime$ meson mass from topological charge density correlator
  in QCD}.
\newblock {\em Phys. Rev.}, D92(11):111501, 2015.

\bibitem{Luscher:2010iy}
M.~Lüscher.
\newblock {Properties and uses of the Wilson flow in lattice QCD}.
\newblock {\em JHEP}, 08:071, 2010.
\newblock [Erratum: JHEP03,092(2014)].

\bibitem{Horvatic:2018ztu}
D.~Horvatić, D.~Kekez, and D.~Klabučar.
\newblock {$\eta'$ and $\eta$ mesons at high T when the $U_A$(1) and chiral
  symmetry breaking are tied}.
\newblock {\em Phys. Rev.}, D99(1):014007, 2019.

\bibitem{Nicola:2018vug}
A.~Gómez~Nicola and J.~Ruiz De~Elvira.
\newblock {Chiral and $U(1)_A$ restoration for the scalar and pseudoscalar
  meson nonets}.
\newblock {\em Phys. Rev.}, D98(1):014020, 2018.

\bibitem{Nicola:2019ohb}
A.~Gómez~Nicola, J.~Ruiz De~Elvira, and A.~Vioque-Rodríguez.
\newblock {The QCD topological charge and its thermal dependence: the role of
  the $\eta'$}.
\newblock {\em JHEP}, 11:086, 2019.

\bibitem{Ishii:2016dln}
M.~Ishii, H.~Kouno, and M.~Yahiro.
\newblock {Model prediction for temperature dependence of meson pole masses
  from lattice QCD results on meson screening masses}.
\newblock {\em Phys. Rev.}, D95(11):114022, 2017.

\bibitem{Mitter:2013fxa}
M.~Mitter and B.-J. Schaefer.
\newblock {Fluctuations and the axial anomaly with three quark flavors}.
\newblock {\em Phys. Rev.}, D89(5):054027, 2014.

\bibitem{Bhattacharya:2014ara}
T.~Bhattacharya et~al.
\newblock {QCD Phase Transition with Chiral Quarks and Physical Quark Masses}.
\newblock {\em Phys. Rev. Lett.}, 113(8):082001, 2014.

\bibitem{Frezzotti:2000nk}
R.~Frezzotti et~al.
\newblock {Lattice QCD with a chirally twisted mass term}.
\newblock {\em JHEP}, 08:058, 2001.

\bibitem{Alexandrou:2016izb}
C.~Alexandrou et~al.
\newblock {Adaptive Aggregation-based Domain Decomposition Multigrid for
  Twisted Mass Fermions}.
\newblock {\em Phys. Rev.}, D94(11):114509, 2016.

\bibitem{Hasenbusch:2002ai}
M.~Hasenbusch and K.~Jansen.
\newblock {Speeding up lattice QCD simulations with clover improved Wilson
  fermions}.
\newblock {\em Nucl. Phys.}, B659:299--320, 2003.

\bibitem{Alexandrou:2018sjm}
C.~Alexandrou et~al.
\newblock {Proton and neutron electromagnetic form factors from lattice QCD}.
\newblock {\em Phys. Rev.}, D100(1):014509, 2019.

\bibitem{Burger:2018fvb}
F.~Burger et~al.
\newblock {Chiral observables and topology in hot QCD with two families of
  quarks}.
\newblock {\em Phys. Rev.}, D98(9):094501, 2018.

\bibitem{Borsanyi:2010bp}
Sz. Borsanyi et~al.
\newblock {Is there still any T\_c mystery in lattice QCD? Results with
  physical masses in the continuum limit III}.
\newblock {\em JHEP}, 09:073, 2010.

\bibitem{Aarts:2019hrg}
G.~Aarts et~al.
\newblock {Spectral quantities in thermal QCD: a progress report from the
  FASTSUM collaboration}.
\newblock In {\em {37th International Symposium on Lattice Field Theory}},
  2019.

\bibitem{progress}
A.~Yu. Kotov, M.~P. Lombardo, and A.~M. Trunin.
\newblock In preparation.

\end{thebibliography}
\end{document}